\documentclass[twocolumn]{aastex63}

\newcommand{\Msun}{\,\ensuremath{\mathrm{M}_\odot}}

\received{June 1, 2019}
\revised{January 10, 2019}
\accepted{\today}
\submitjournal{ApJ}

\shorttitle{{\it r}-process in globular clusters}
\shortauthors{Tarumi et al.}
\graphicspath{{./}{figures/}}

\begin{document}

\title{Internal {\it r}-process abundance spread of M15 
and a single stellar population model}

\correspondingauthor{Yuta Tarumi}
\email{yuta.tarumi@phys.s.u-tokyo.ac.jp}

\author{Yuta Tarumi}
\affiliation{Department of Physics, School of Science, The University of Tokyo, Bunkyo, Tokyo 113-0033, Japan}

\author[0000-0001-7925-238X]{Naoki Yoshida}
\affiliation{Department of Physics, School of Science, The University of Tokyo, Bunkyo, Tokyo 113-0033, Japan}
\affiliation{Kavli Institute for the Physics and Mathematics of the Universe (WPI), UTIAS, The University of Tokyo, Chiba, 277-8583, Japan}
\affiliation{Research Center for the Early Universe, School of Science, The University of Tokyo, Tokyo, 113-0033, Japan}

\author{Shigeki Inoue}
\affiliation{Center for Computational Sciences, University of Tsukuba, Ten-nodai, 1-1-1 Tsukuba, Ibaraki 305-8577, Japan}
\affiliation{Chile Observatory, National Astronomical Observatory of Japan, Mitaka, Tokyo 181-8588, Japan}
\affiliation{Department of Cosmosciences, Graduates School of Science, Hokkaido University, Sapporo, Hokkaido 060-0810, Japan}



\begin{abstract}

The member stars in globular cluster M15 show a 
substantial spread in the abundances of $r$-process elements. 
We argue that a rare and prolific $r$-process event enriched the natal cloud of M15 in an inhomogeneous manner.
To critically examine the possibility, we perform cosmological galaxy formation simulations and study the physical conditions for the
inhomogeneous enrichment.
We explore a large parameter space of the merger event 
time and the site.
Our simulations reproduce the large $r$-process abundance spread if a neutron-star merger occurs at $\sim 100$\,pc away from the formation site of the cluster and in a limited time range of a few tens million years before the formation.
Interestingly, a bimodal feature is found in the Eu abundance distribution in some cases, similarly to that inferred from recent observations.
M15 member stars do not show clear correlation between the abundances of Eu and light elements such as Na that
is expected in models with two stellar populations. We thus argue that a majority of the stars in M15 are formed in a single burst.
The ratio of heavy to light $r$-process element abundance $\mbox{[Eu/Y]} \sim 1.0$ is consistent with that of the so-called $r$-II stars,
suggesting that a lanthanide-rich $r$-process event dominantly enriched M15.

\end{abstract}

%


\section{Introduction}

Globular clusters (GCs) are compact, nearly spherical clusters with old stellar populations. Typically, more than $10^{5}$ stars gather within $\sim 10$\,pc, making the most densely populated environments in the Universe. 
It was thought that stars in a GC are of single population: all the stars are formed at once, and share the same stellar age and chemical abundance. Recent studies find substantial scatter in the color-magnitude diagram (CMD)
(see \citealt{2008Renzini_GC_MP} and references therein), and the spread is confirmed in almost all GCs older than 2\,Gyr. It appears that GCs host multiple populations commonly \citep{2018Bastian}.

Peculiar features are found in the chemical abundance patterns
of GC stars. The abundances of light metals such as sodium (Na) and oxygen (O) show prominent spread within each cluster. Moreover, the abundances clearly (anti-)correlate with each other, suggesting that high-temperature hydrogen-burning have played a role. This is in stark contrast with the nearly uniform iron (Fe) abundance in GCs\footnote{An exception is $\omega$ centauri, which shows internal metallicity spread (e.g. \citealt{2010JohnsonPilachowski_OmegaCen}).}.

Elements heavier than Fe are also found in GCs and the abundances are measured. The heavy elements are mainly synthesized in two neutron-capture processes: the $s$- and $r$-process \citep{1957BBFH}. The dominant  $s$-process factory is thought to be asymptotic-giant branch (AGB) stars \citep{2011Kappeler_sprocess_review}, but astrophysical sites for the $r$-process are still under debate, although recent multi-messenger observations of the neutron-star merger (NSM) renders it the most promising candidate (GW170817, see, e.g.  \citealt{2017Cowperthwaite_NSM, 2018Hotokezaka_rprocess_by_NSM, 2021Cowan_RprocessReview}).

M15 is a unique GC with unquestionable spread in europium (Eu) abundance \citep{1997Sneden, 2006Otsuki, 2011Sobeck, Roederer11_GCrprocess, 2011Cohen, 2013Worley}. For most GCs, the abundances of Eu is consistent with being uniform within observational uncertainty. 
The physical process that caused the abundance spread is not well understood.
An interesting possibility is $r$-process enrichment between multiple star formation epochs \citep{2017BekkiTsujimoto_GC, 2019Zevin_GC}. Since a GC is thought to host multiple star formation epochs, $r$-process event after the first-generation (FG) star formation can enrich only the next generation stars.

In the present paper, 
we consider inhomogeneous mixing of elements in the natal cloud as proposed in \citet{Roederer11_GCrprocess}. A merger of wandering binary neutron-stars (BNS) synthesize and deposit the $r$-process elements in a small portion of gas in the star-forming region. The ejecta is mixed within the interstellar medium (ISM) mainly by turbulence, but the mixing process takes as long as a few to several tens million years.
GC stars formed in such an inhomogeneous cloud can naturally have elemental abundance variations.
In Section 2, we describe our simulation method.
Formation of the target host galaxy is described in Section 3. Then we show the results in Section 4. In Section 5, we discuss the implications of our result. 

\section{Method}

We study the physical conditions
under which a GC can have Eu abundance spread. To this end, we first run a cosmological simulation and identify compact small clusters (GCs). 
We then perform re-simulations of the GC formation 
by injecting $r$-process elements into the gas cells near the formation site. We study the abundances of $r$-process elements in the formed stars.

\subsection{General settings}

\subsubsection{cosmologial simulations}

We use a moving-mesh hydrodynamic simulation code \textsc{arepo} \citep{Springel10_AREPO, Pakmor16_AREPO_improvement, 2019AREPOrelease}. We adopt the Planck 2018 cosmological parameters \citep{Planck2018}: $\Omega_{m} = 0.315, \Omega_{b} = 0.049, \sigma_{8} = 0.810, n_{s} = 0.965, H_{0} = 67.4\ \mathrm{km\,s^{-1}\,Mpc^{-1}}$. The initial conditions are generated with \textsc{music} \citep{Hahn11_MUSIC}. First, we run a low-resolution simulation with a cubic box of 10\,cMpc $h^{-1}$ on a side. Then we identify a dark matter halo of $2.5\times 10^{11} \Msun$ at redshift 4,
which is selected as a host of GCs. 
Next, we run a zoom-in simulation with increasing the mass resolution so that the mass of a dark matter particle is $1.0\times 10^{5}\,\Msun$ and the typical mass of each gas cell is $1.9\times 10^{4}\,\Msun$. With this mass resolution, a typical target GC with 
stellar mass of $4\times 10^{5} \Msun$ is resolved by more than $20$ gas elements, which allows us to represent spatial variation of Eu abundance.

\subsubsection{Star formation and feedback}
A gas cell bears star particles when the hydrogen number density of the gas is higher than $10^{4}$ per 1 cm$^{3}$. Star formation is allowed only in cool ($T < 10000$ K) gas cells. This temperature condition is imposed to avoid star-formation in dense gas shocked by supernovae (SNe).

We adopt a momentum injection method for the SN feedback as in \citet{2019Marinacci}.
Briefly, if the mass resolution of a  simulation is not as good as $\sim 1000$ \Msun, the momentum enhancement during the sedov-taylor phase is taken into account \citep{2014kimm, 2018Hopkins_FIRE, 2019Marinacci}. Taking $P_\mathrm{SN} = \sqrt{2E_\mathrm{SN}M_\mathrm{SN}}$ as  default momentum, we boost it by a factor $f_\mathrm{enhance} = \sqrt{1+M_\mathrm{cell}/\Delta M}$. Here the $E_\mathrm{SN}$ and $M_\mathrm{SN}$ are energy of supernova and mass of supernova ejecta, which we assume as $10^{51}\,\mathrm{erg}$ and $15.2\,\Msun$, and $\Delta M$ is the mass imparted to a gas cell. There is a maximum in the boosted momentum $p_\mathrm{terminal}$ from each SN \citep{1988Cioffi}:
\begin{equation}
    p_\mathrm{terminal} = 4.8\times 10^{5} E_\mathrm{SN}\biggl(\frac{n_\mathrm{H}}{1\ \mathrm{cm}^{-3}}\biggr)^{-1/7}f(Z)^{3/2} \Msun\ \mathrm{km\ s^{-1}},
\end{equation}
where $f(Z)=\mathrm{min}[(Z/Z_{\odot})^{-0.14}, 2]$.

The importance of radiation feedback has been pointed out in previous studies \citep{2013Stinson}. We implement a simple model for radiation feedback. We assume that each star particle keeps injecting energy for the first 4\,Myr after formation. The energy emission rate is $2.3\times 10^{36}\ \mathrm{erg\ s^{-1}}$ per solar-mass. This rate corresponds to the number of UV photons of $8.0\times 10^{46}\ \mathrm{s^{-1}}$ with an average energy of $\langle h\nu \rangle = 18\ \mathrm{eV}$ (e.g. \citealt{2020Fukushima_GCformation}). We turn off radiative cooling for 4\,Myr in cells affected by the radiation feedback.

The kernel function for both the radiation feedback and the SNe feedback is proportional to the solid angle subtended by the cell viewed from the SN site (star particle). Namely,
\begin{equation}
    w_i = \frac{\Delta\Omega_{i}}{4\pi}.
\end{equation}
The kernel is calculated for the closest 64 cells around the star particle.

\begin{figure*}[htbp]
    \centering
	\includegraphics[width=\columnwidth]{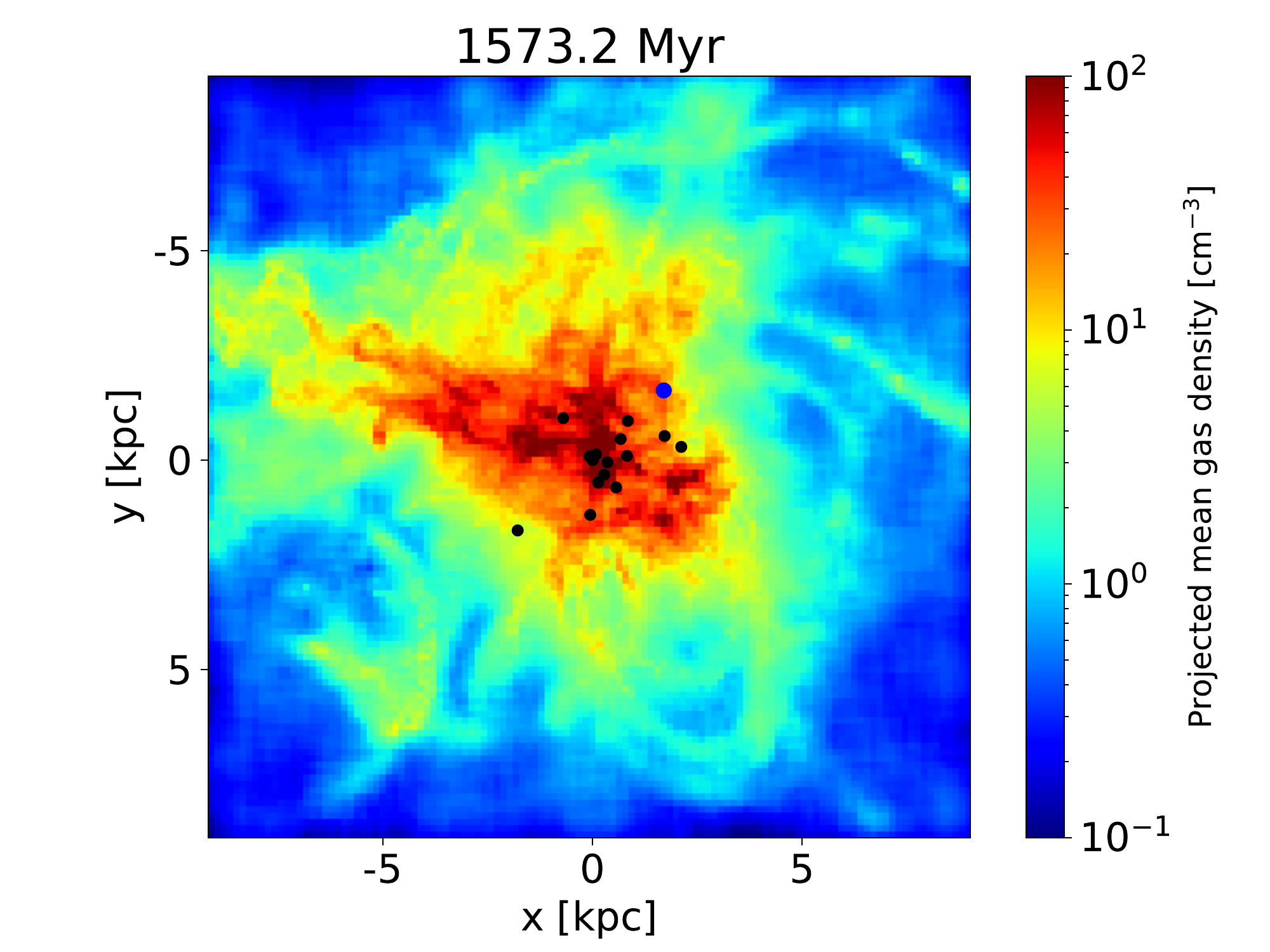}
	\includegraphics[width=\columnwidth]{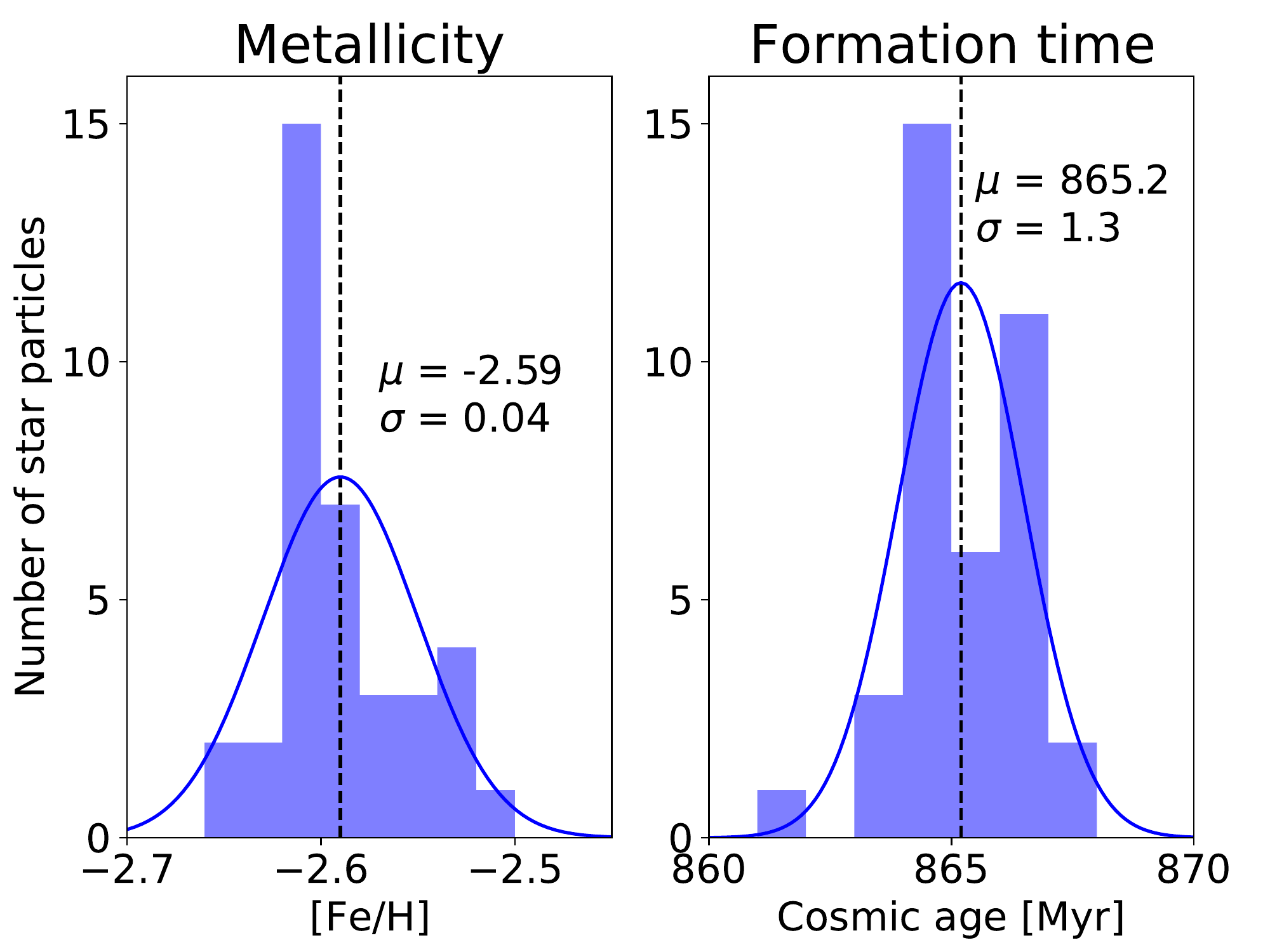}
    \caption{Left panel: mean projected density of gas of the simulated galaxy at cosmic time $\sim 1.6\,\mathrm{Gyr}$ (redshift 4.0). The positions of the GCs are marked with black, and the one we analyze in this paper is marked with blue. Right panel: Histograms of metallicity and formation times of star particles in the target GC.}
    \label{fig:snapshots}
\end{figure*}

\subsubsection{Identification of GCs}
GCs are typically very compact and self-bound objects with an approximately constant [Fe/H]. We identify candidate star clusters as follows. First, we locate star particles in a high stellar density environment with at least $3\times 10^{5}\,\Msun$ of stars within 10\,pc.
Next, we group the stars using the k-means clustering algorithm. We obtain $19$ star particle groups that are sufficiently dense and well separated from each other. Two of them are very massive ($>10^{8}\,\Msun$) and contain star particles with various ages and metallicities. These are likely nuclear clusters in centers of galaxies, and we eliminate them from our sample. The other 17 clusters have had short ($\lesssim 10\,\mathrm{Myr}$) star formation duration and have homogeneous iron abundances. The range of stellar masses of the 17 GCs is $4.0\times 10^5 \Msun$ -- $2.8\times 10^{6} \Msun$.
Finally, we regard stars within 89\,pc (300\,cpc $h^{-1}$) from the cluster center as members of the GC. We check if the clusters survive in the final snapshot at $z = 4$. These star clusters likely remain bound until the present epoch, 
to be observed as ``old'' GCs. We note that, if we change the distance criterion to 5\,pc and 20\,pc, the number of the identified GC varies to 10 and 21.

Figure~\ref{fig:snapshots} shows the projected gas density at redshift 4. 
Among the simulated GCs, we select one as our target GC, which is marked by a large blue circle. Any of these GCs would give similar results and we pick up this sample as a typical example.
The positions of other GCs are shown as black filled circles. 
The right panel of Figure~\ref{fig:snapshots} shows the distributions of the formation 
times and metallicities  of the member stars of our target GC. The first star particle is formed at a cosmic age of 861.4\,Myr, and star formation continues for 5.8\,Myr. The mean formation time is 865.2\,Myr, which corresponds to $z=6.5$. The mean of iron abundance [Fe/H] is $-2.59$, and its standard deviation is 0.04. The total stellar mass of the target GC at our final snapshot is $6.4\times 10^{5}$ \Msun. The small Fe abundance spread is consistent with that of a GC in observations ($\sim 0.06$\,dex, see \citealt{1997Carretta_homogeneous_Fe}). 

\subsection{Numerical experiments}
We test our inhomogeneous mixing model 
by using direct numerical simulations.
We take an approach similar to our previous paper \citep{2020Tarumi}, where we model an $r$-process event, NSM, as a point explosion. We deposit Eu to the interstellar medium (ISM) that is going to form the target GC. We follow the dispersal and dilution of Eu to the surrounding ISM. The main stellar population of the GC is formed after the $r$-process element enrichment. 

We use 20 snapshots dumped at times before the GC formation, separated by $3-5$\,Myr from the adjacent output time. On each snapshot, we first determine the formation site of the target GC.
To this end, we first identify the IDs of gas cells at a moment when the GC is forming. We then look for the cells back in time. In each snapshot we identify the flagged gas cells, and determine the GC formation planned site as the median of all the flagged gas cells.
We then deposit Eu into the gas cells around the planned site.
To investigate the relation between the distance and the resulting Eu abundance spread, we also test off-center NSM. First we take four concentric spheres centered at the planned site. On each surface, we pick up 26 points, which are $(\theta, \phi) = (0, 0), (\pi/4, i), (\pi/2, i), (3\pi/4, i), (\pi, 0)$ with $i$ runs every $\pi/4$ from 0 to $7\pi/4$ in spherical polar coordinates. In total, we take $26\times 4 = 104$ points in addition to the planned site as the explosion sites. The radii of the spheres are chosen considering the time duration between the element deposition and the GC formation. Let $t$ be the deposition time before the GC formation. For snapshots with ($t < 30$\,Myr, $30\,\mathrm{Myr} < t < 60\,\mathrm{Myr}$, $60\,\mathrm{Myr} < t$), the radii are (100, 200, 300, 400)\,pc, (200, 400, 600, 800)\,pc, and (300, 600, 900, 1200)\,pc, respectively. 

We inject a scalar component that mimics Eu into gas cells around the explosion. 
Eu is injected into a gas cell if it is within the snowplow radius $r_\mathrm{sp}$ of an injection center. The snowplow radius $r_\mathrm{sp}$ is calculated as follows \citep{GalaxyFormaitonAndEvolution}. At the end of the self-similar Sedov-Taylor phase, the blast-wave radius is 
\begin{equation}
    r_\mathrm{sh} = 23\times \biggl(\frac{n}{1\mathrm{cm^{-3}}}\biggr)^{-19/45}\times \biggl(\frac{E}{10^{51}\mathrm{erg}}\biggr)^{13/45}\,\mathrm{pc},
\end{equation}
and the velocity is
\begin{equation}
    v_\mathrm{sh} = 200\times \biggl(\frac{n}{1\mathrm{cm^{-3}}}\biggr)^{2/15}\times \biggl(\frac{E}{10^{51}\mathrm{erg}}\biggr)^{1/45} \mathrm{km~s^{-1}},
\end{equation}
where $n$ is the number density in a unit of cm$^{-3}$. After that, the shell expands conserving the momentum (snowplow phase). When the shell expansion velocity becomes comparable to the turbulent velocity of the ISM ($\sim 10$ km s$^{-1}$), the snowplow phase ends and it blends into the ISM. The radius at the end of the snowplow phase is 
\begin{equation}
    r_\mathrm{sp} = r_{\rm sh}\times \biggl(\frac{v_{\rm sh}}{10\mathrm{km~s^{-1}}}\biggr)^{1/3}.
\end{equation}
This approach typically distributes Eu to $\sim 10^{5} \Msun$ of gas cells.

The amount of injected Eu is $5\times 10^{-5}$ \Msun, which is consistent with GW170817 \citep{2017Cowperthwaite_NSM} and with estimates from Reticulum II observation \citep{2016AlexJi_Nature}. The amount of Eu assigned to each gas cell is set proportional to the volume of the gas cell. 

Observations suggest that the $r$-process elements in M15 is produced with more than one event, with different ratio of light-to-heavy elements \citep{2006Otsuki}. We assume two kinds of $r$-process events, lanthanide-poor and lanthanide-rich, exist as the observations suggest. The natal cloud is enriched with both of these events, and it has an abundance of $\mbox{[Eu/H]} = -2.1$. This value is motivated by the fact that the metallicity of M15 is $\mbox{[Fe/H]} \sim -2.4$ and Milky Way's halo stars with similar metallicity have $\mbox{[Eu/Fe]} \sim 0.3$. 
This constant ``pre-enrichment'', although we do not specify its origin, makes it possible to directly compare our simulation results with M15 observations. Further discussion on the origin of first-peak $r$-process elements and lanthanides is presented in Section 4.3.

\section{Results}

\begin{figure*}[t]
	\centering
	\includegraphics[width=\columnwidth]{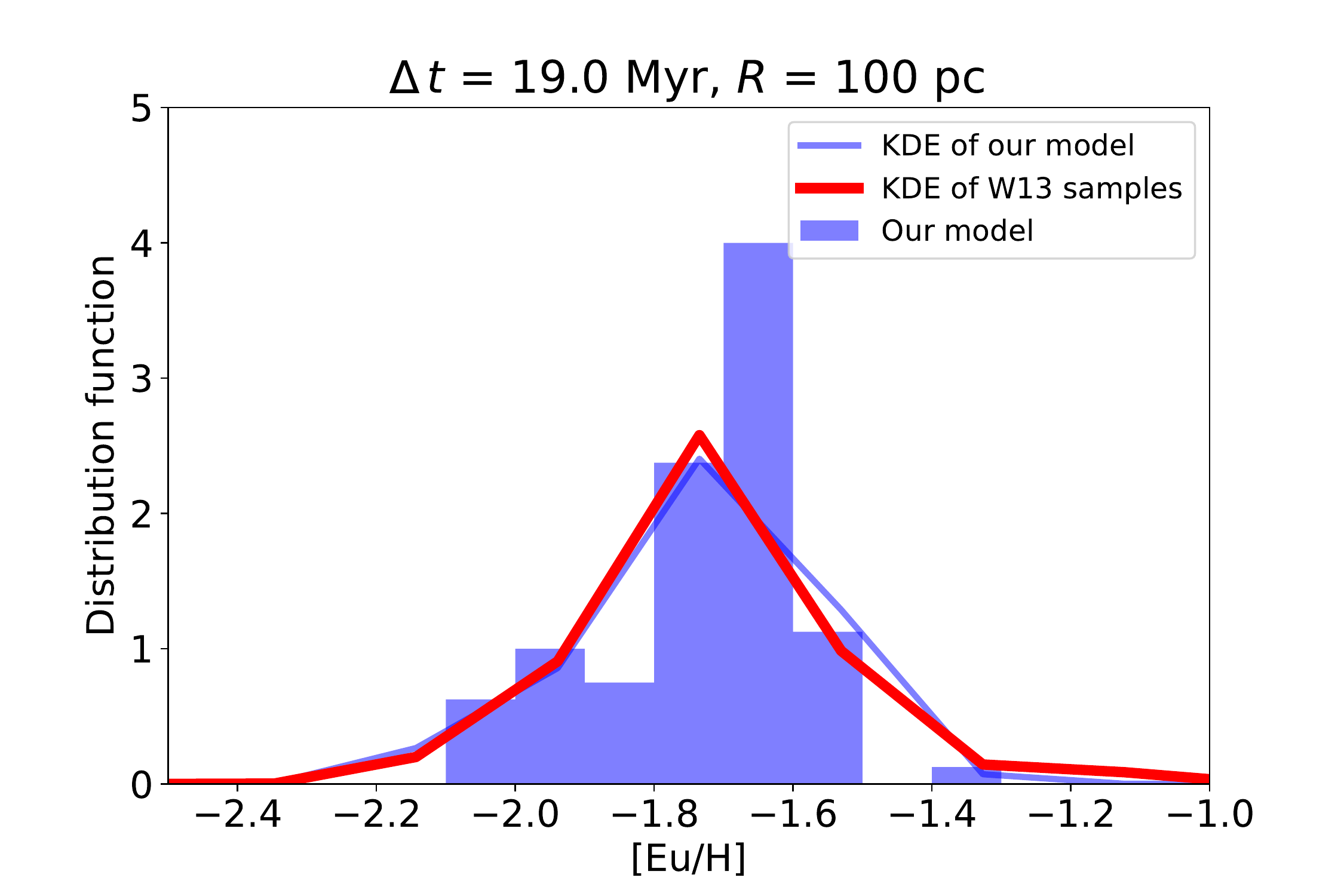}
	\includegraphics[width=\columnwidth]{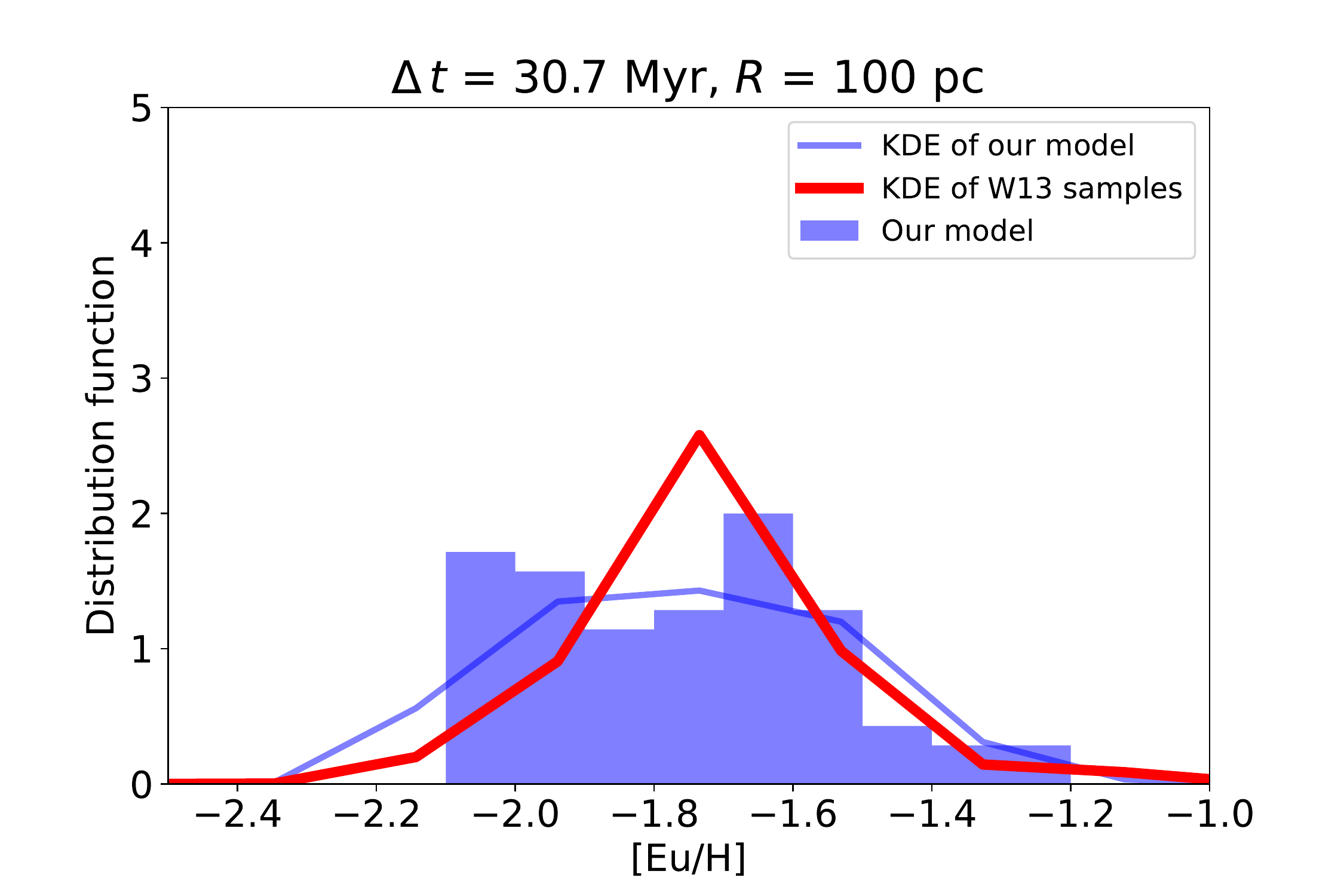}
	\includegraphics[width=\columnwidth]{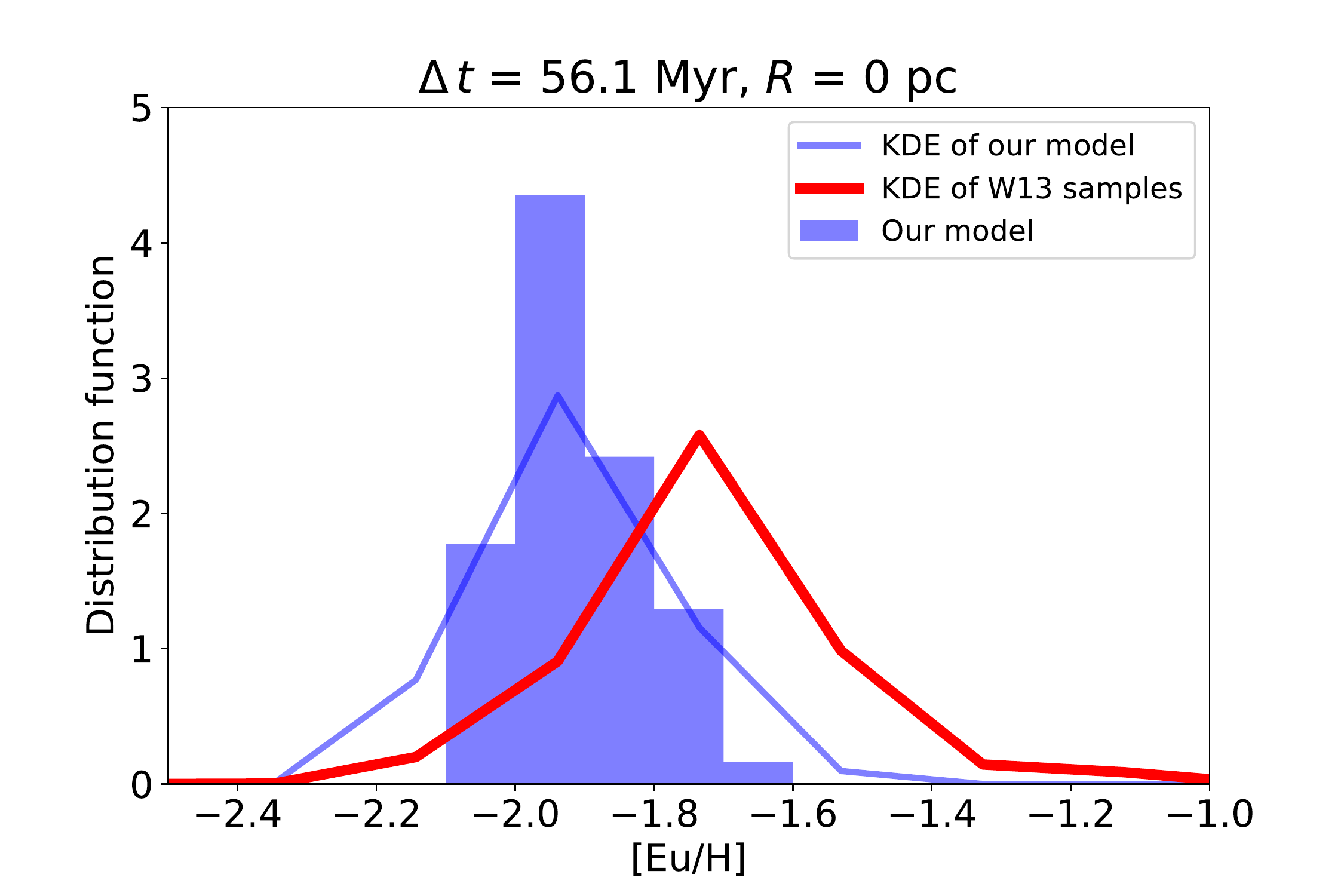}
	\includegraphics[width=\columnwidth]{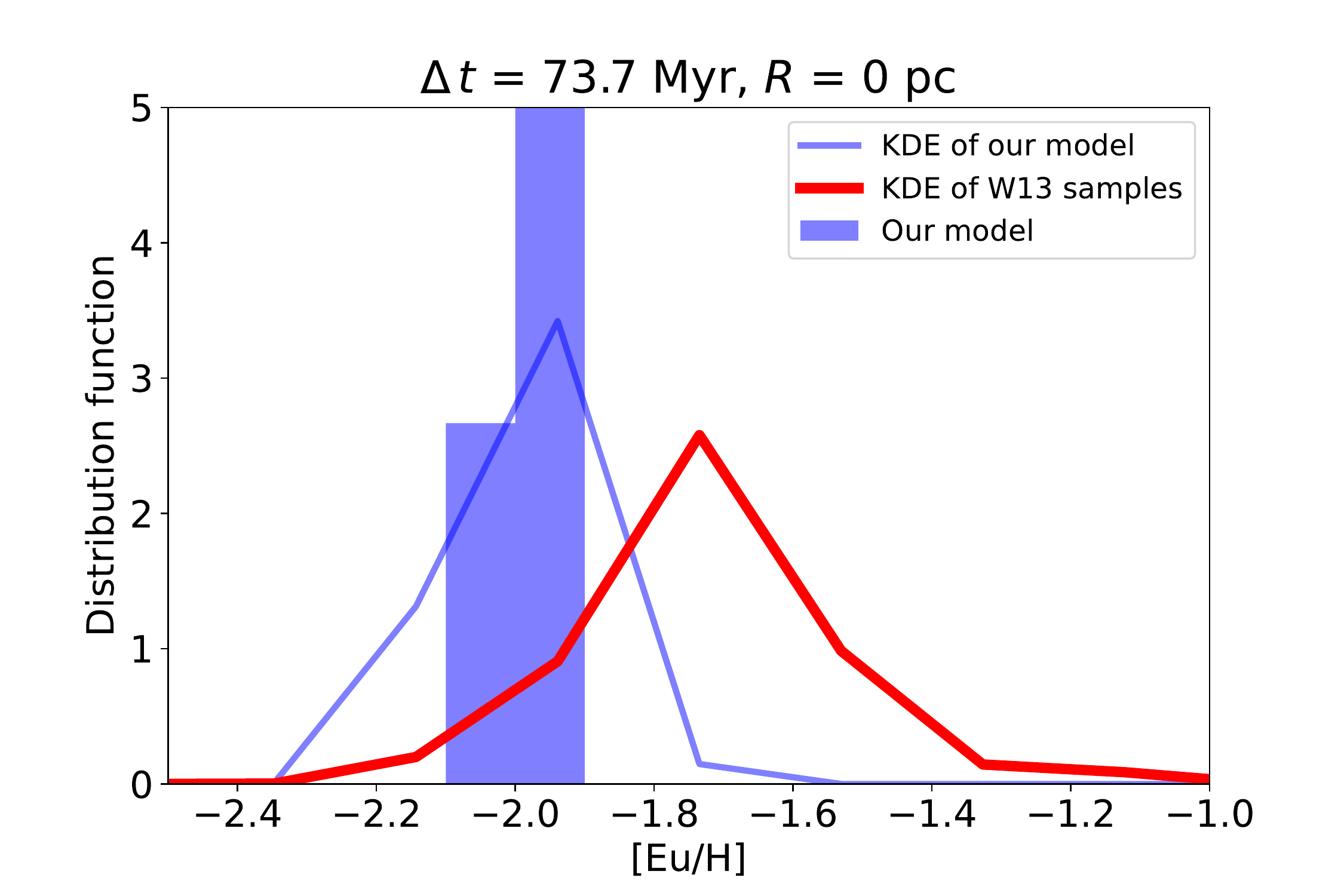}
    \caption{Europium abundance distribution of M15 stars and the simulated GC. The red line shows the kernel density estimation (KDE) of stars in M15 in \citet{2013Worley}, the blue histograms are the results of our simulations, and the blue lines are the KDE of our simulation results. The bandwidth is 0.1 dex. The top two panels are ones of (19.0\,Myr, 100\,pc) and (30.7\,Myr, 100\,pc) models that reproduce the [Eu/H] distribution well. These distributions have the standard deviations of $\sigma > 0.15$\,dex. The bottom panels are ones of (56.1\,Myr, 0\,pc) and (73.7\,Myr, 0\,pc) models that still show inhomogeneity, although their standard deviations are smaller than or comparable to the observed spread, i.e., $\sigma < 0.15$\,dex.}
    \label{fig:parametric Eu/H}
\end{figure*}

\subsection{Eu abundance distribution}

In Figure~\ref{fig:parametric Eu/H}, we show Eu abundance distributions for four different snapshots. The red lines in the four panels show the kernel density estimation of Eu abundance distribution for M15 stars \citep{2013Worley}. The blue histograms are the results of our numerical simulations. The top left panels show the result of one of our ``best'' models in which Eu is deposited 19.0\,Myr before the GC formation. The explosion site is 100\,pc away from the formation site. The observed Eu abundance distribution is reproduced well. 
Other realizations with the same model parameters with $\Delta t$ = 19.0\,Myr and $R$ = 100\,pc have comparable Eu abundance scatter, although with somewhat different distributions. 
The top right panel shows a slightly different model with $\Delta  t$ = 30.7\,Myr and $R$ = 100\,pc. Although the shape of the distribution function is different, it still has a significant abundance spread within one system. Interestingly, this particular realization yields an apparent double-peak distribution. The low-abundance peak is at [Eu/H] = $-2.1$; some stars are barely enriched by the NSM. It is interesting that 
the apparent bimodality can be reproduced 
by our inhomogeneous ISM model without invoking multiple stellar formation epochs. 

The bottom two panels of Figure~\ref{fig:parametric Eu/H}
are results for models that do not reproduce the observed distribution. 
In the model with $\Delta t$ = 56.1\,Myr and $R$ = 0\,pc (bottom left), 
even if an NSM occurs at the center, the ejecta dilutes with a large amount of ISM. 
We find a small Eu abundance spread with the standard deviation of 0.1\,dex, which is comparable to observational uncertainties. 
Another model is with $\Delta t$ = 73.7\,Myr and $R$ = 0\,pc (bottom right). In this long duration model, Eu from the last NSM is almost completely mixed within the ISM, and the overall Eu abundance and its scatter are inconsistent with observation.
Based on the simulation results, we argue that the observed Eu abundance can be reproduced if the $r$-process enrichment 
(NSM) occurs in the outerpart of the natal cloud, and if a short period of time is available for ISM mixing.

\subsection{Distribution of Eu abundance spread}
\label{sec:probability}

\begin{figure}
	\includegraphics[width=\columnwidth]{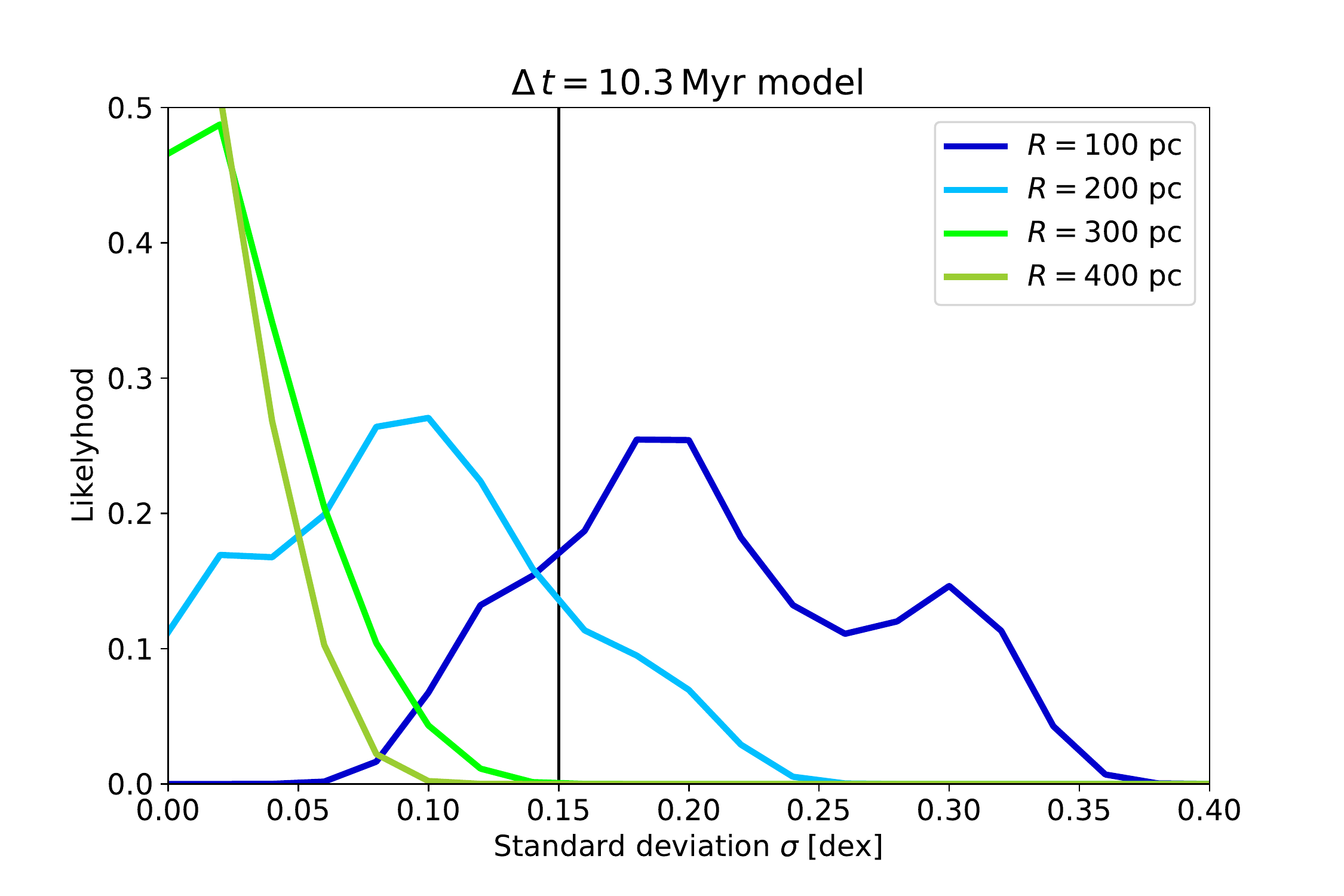}
	\includegraphics[width=\columnwidth]{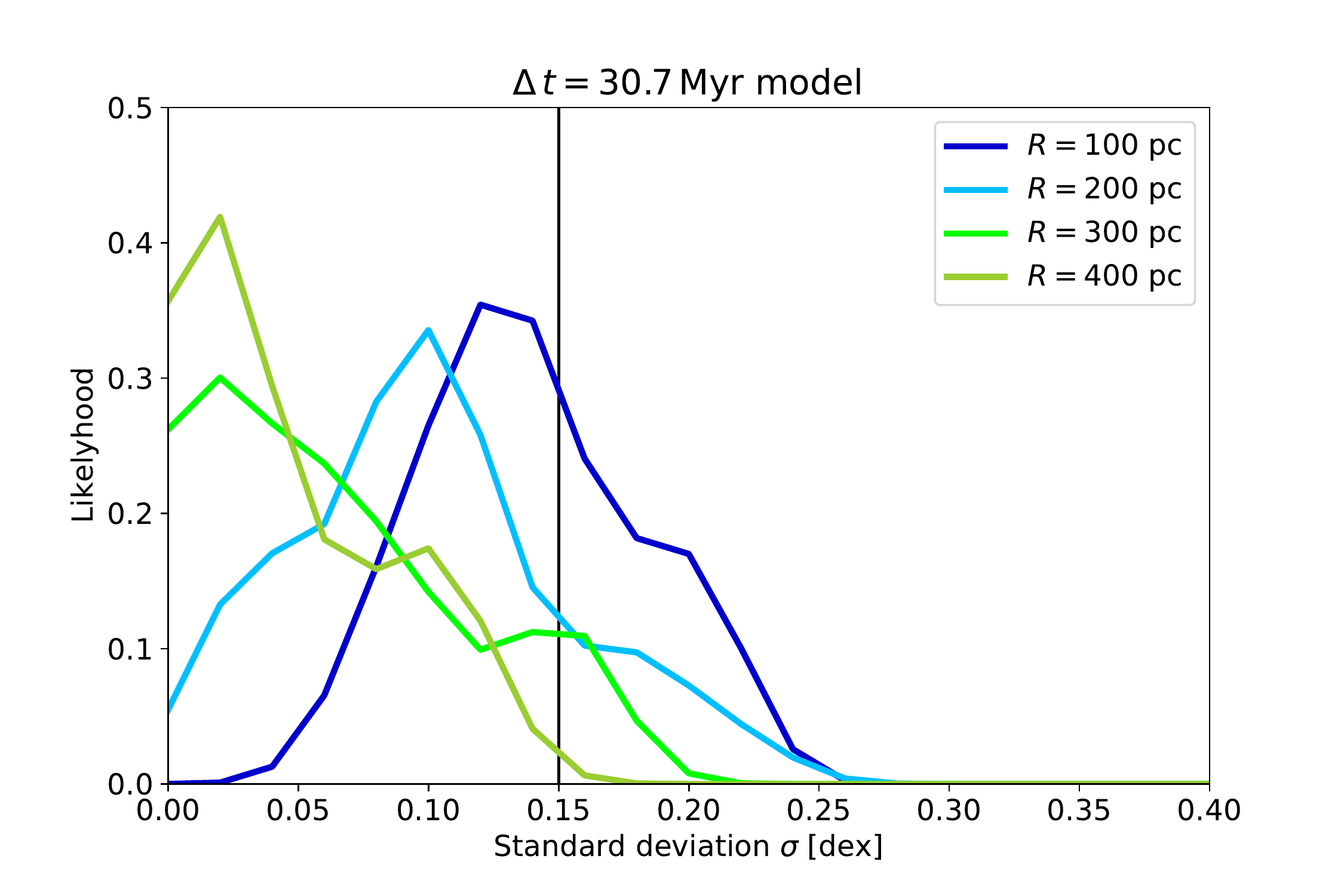}
    \caption{
    Distribution of the standard deviation of Eu abundance [Eu/H] of GCs. The vertical line at $\sigma = 0.15\,\mathrm{dex}$ denotes the observed distribution spread of M15 \citep{2013Worley}. We use kernel density estimation for the illustration purpose. The bandwidth is 0.02 dex. 
    }
    \label{fig:sigma distribution}
\end{figure}

Figure~\ref{fig:sigma distribution} shows the distribution of Eu abundance spread for $\Delta\,t = 10.3$ and $30.7\,\mathrm{Myr}$ models. In $\Delta\,t = 10.3\,\mathrm{Myr}, R = 100\,\mathrm{pc}$ models, more than 80\,\% show abundance spread comparable to or more than M15. The fraction decreases as we increase the distance, and none of $R \gtrsim 300\,\mathrm{pc}$ models show observable spread. In longer duration models ($\Delta\,t = 30.7\,\mathrm{Myr}$), the inhomogeneous probability for $R = 100\,\mathrm{pc}$ models decrease to 35\,\%. This is because the NSM ejecta is diluted to a larger volume. In this long duration model, however, 15\,\% of $R = 300\,\mathrm{pc}$ models show significant Eu spread. Time evolution decreases the probability for small $R$ models, but gives chances for relatively large $R$ models to show Eu abundance spread.

We can calculate the volume in which an NSM ejecta remains inhomogeneous. If a GC is formed within, it would have Eu abundance spread $\sigma$ of more than $\sigma > 0.15$\,dex. The volume is $\sim 0.02$\,kpc$^{3}$, and it is roughly constant for the first 50\,Myr. This reflects the fact that the probability decrease for low $R$ models and increase for high $R$ models have compensated with each other. After $\sim 50$\,Myr has passed, the volume quickly decreases to zero due to the dilution.

\section{Discussion}

We have identified a successful case 
where GC members stars show substantial spread in
Eu abundance. In this section, we discuss the characteristics
of the stellar population(s) of M15, and propose a viable model
for the formation and evolution.

\subsection{Sodium abundance}

\begin{figure}
    \begin{center}
	\includegraphics[width=\columnwidth]{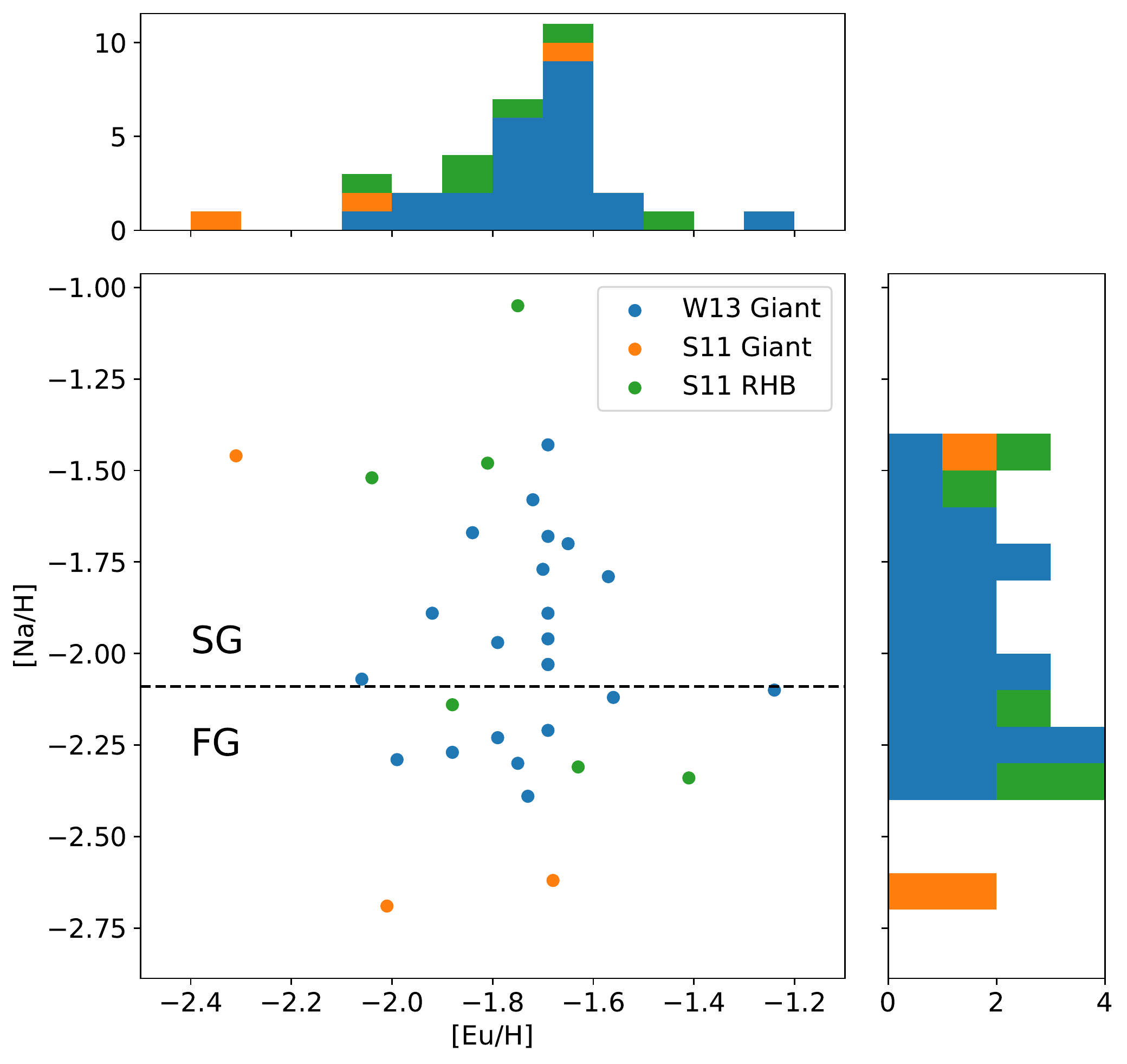}
	\includegraphics[width=0.3\columnwidth]{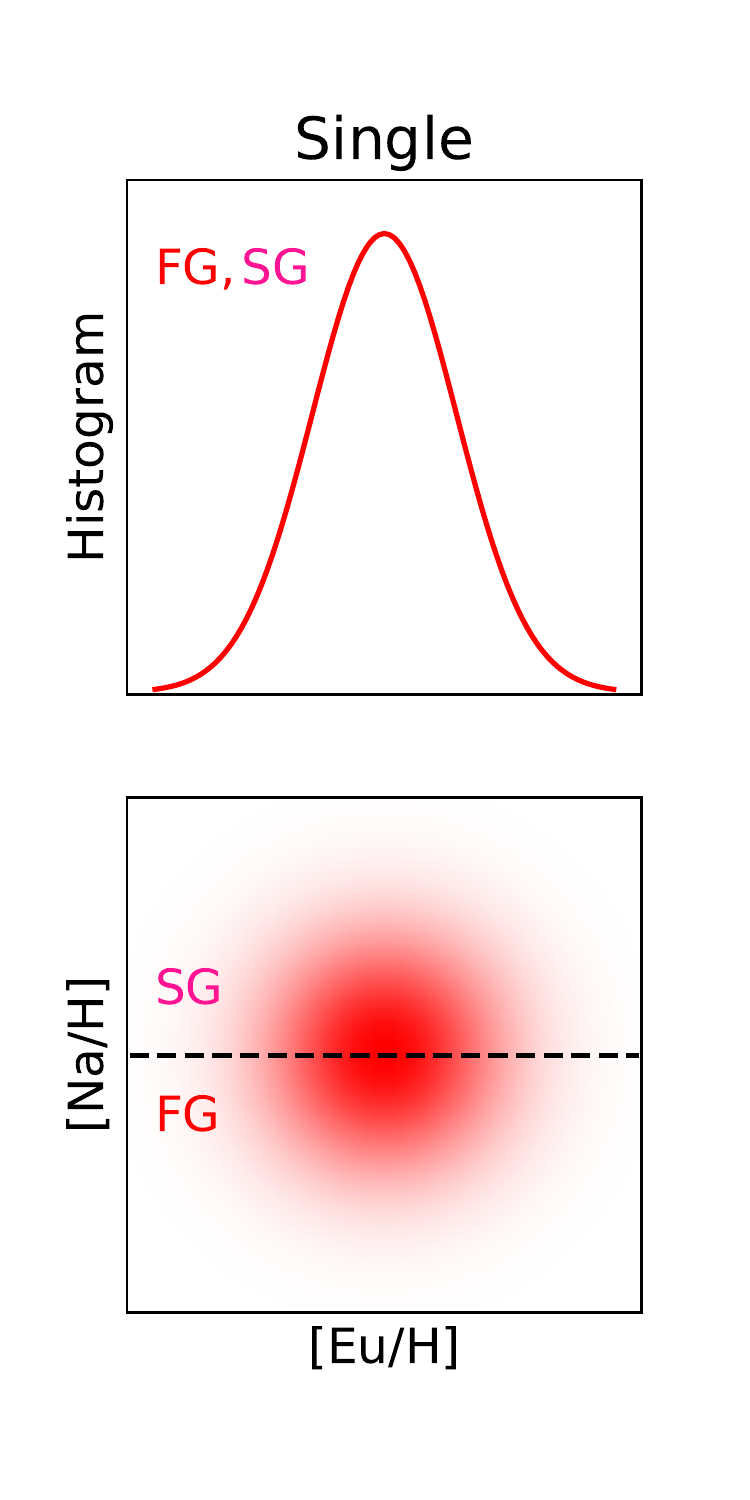}
	\includegraphics[width=0.3\columnwidth]{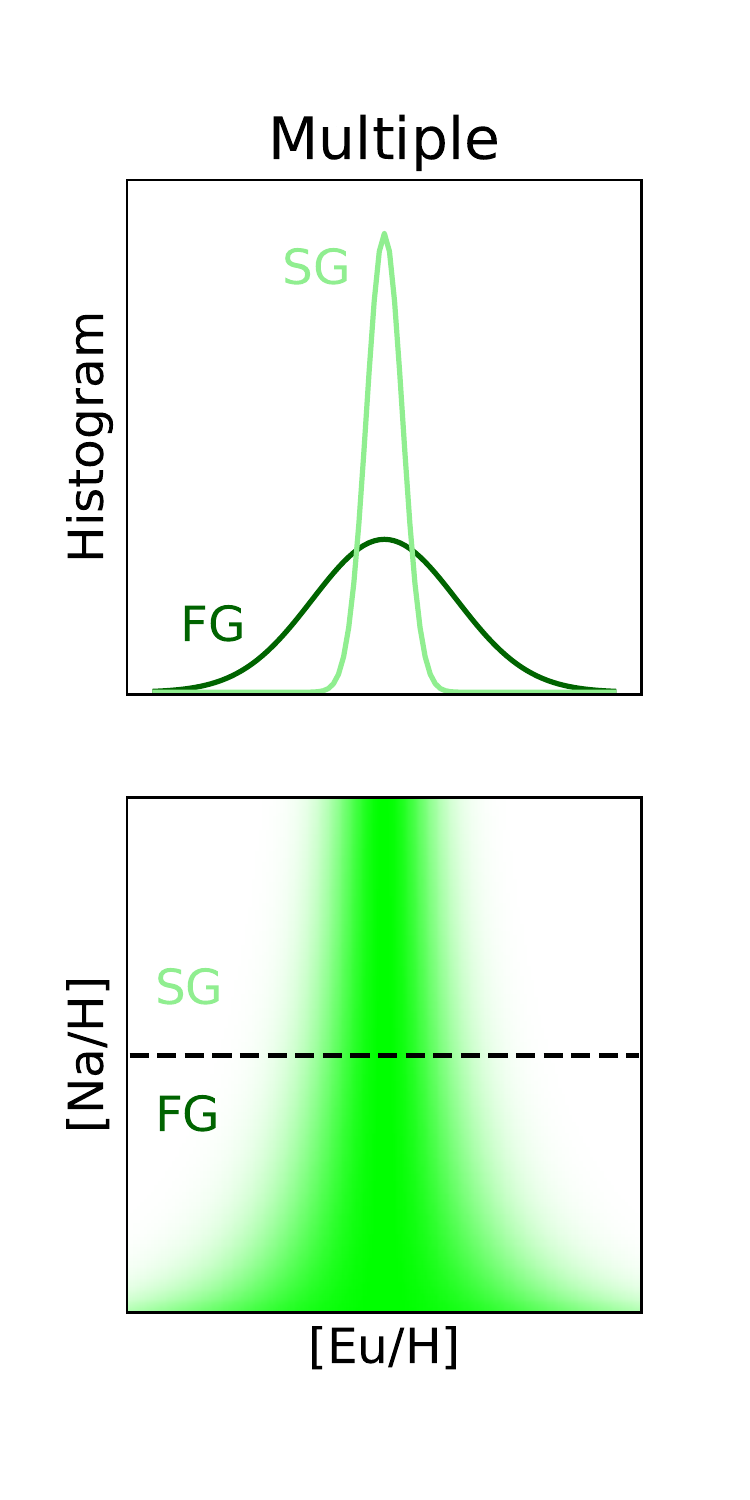}
	\includegraphics[width=0.3\columnwidth]{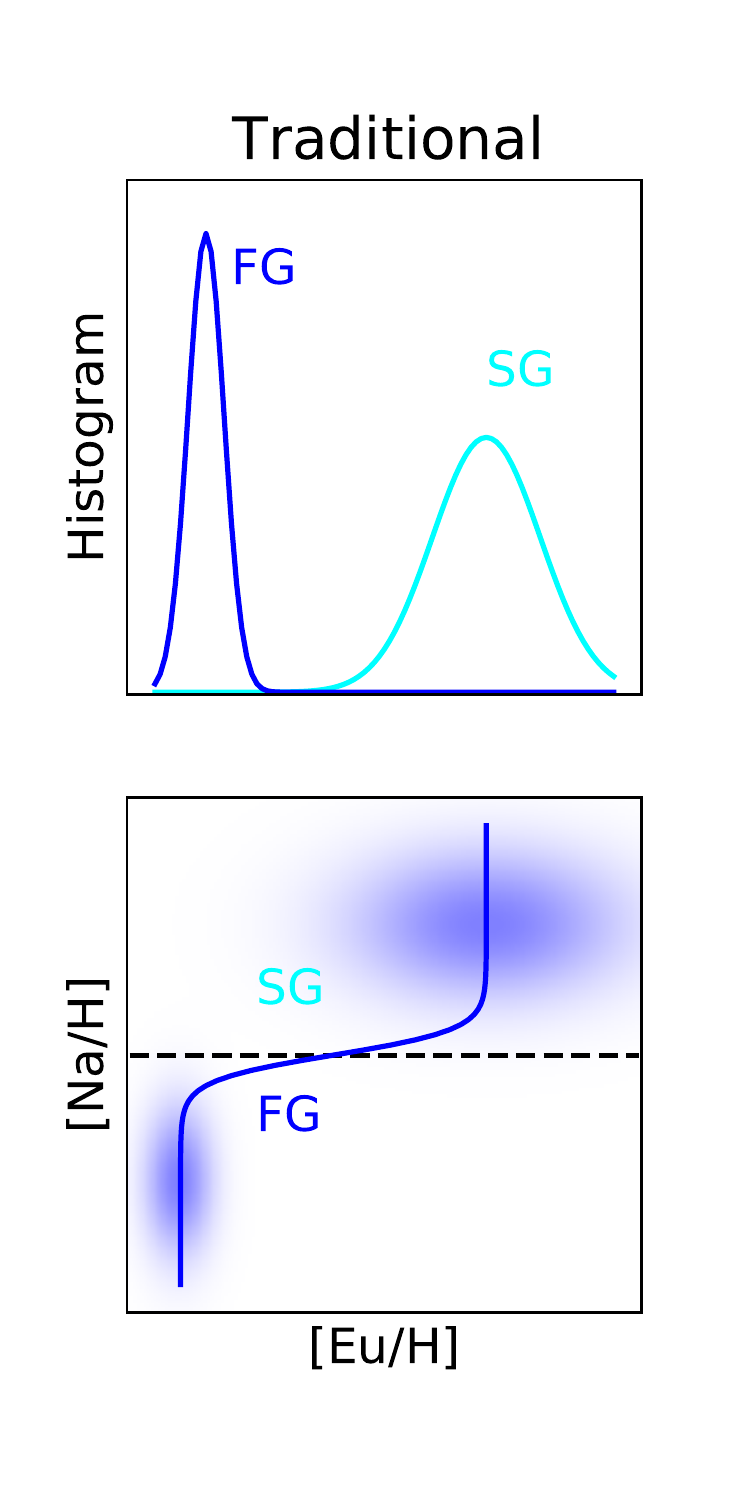}
	\end{center}
    \caption{Top panel: distribution of [Na/H] and [Eu/H] in \citet{2013Worley} and \citet{2011Sobeck}. Two histograms show the distribution of [Eu/H] and [Na/H]. Bottom panels: expected distributions in three different scenarios. Ours are the left two panels, depending on whether the GC stars form in one or multiple bursts.}
    \label{fig:M15_NaEu}
\end{figure}

In Figure~\ref{fig:M15_NaEu}, we show the observed distribution of M15 stars on the [Eu/H] - [Na/H] plane and the corresponding histograms. We see a marginal bimodality in [Eu/H] with two peaks at $\mbox{[Eu/H]} \simeq -2.1$ and $\mbox{[Eu/H]} \simeq -1.7$ 
(see also Section 3.1). We do not find  correlation between Eu and Na. Based on these facts, we discuss a few scenarios for the formation and $r$-process enrichment of M15.

A popular formation model of GCs assumes multiple star formation epochs. During the formation of the first-generation (FG) stars, some BNSs are formed. One (or more) of the BNS merge before the second-generation (SG) star formation and enrich the cluster. Then, the SG stars have higher [Eu/H] than the FG stars. The observed Eu abundance spread can be reproduced in this scenario
\citep{2017BekkiTsujimoto_GC, 2019Zevin_GC}. 
An important feature is that the $r$-process enrichment affects only the SG stars. In the bottom right panel of Figure~\ref{fig:M15_NaEu}, we show schematically the expected distribution function. 
The FG stars have abundances similar to those of halo stars, and we do not expect Eu spread among them. The FG stars are not affected by the $r$-process event nor by the neon-sodium chain (NeNa-chain). On the other hand, the SG stars are highly enriched with both the $r$-process event {\it and} the NeNa-chain, therefore appearing on the top right. For SG stars there could be a spread in Eu abundance, depending on the timescale of star formation, similarly to our model studied in Section 3. In this case, a clear bimodal distribution of [Eu/H] can be seen. Another prediction is the correlation of Eu to Na (and other light elements). It is known that GCs have Na-O anticorrelation (see, e.g. \citealt{2018Bastian}), which is a sign of hydrogen-burning between the FG and the SG star formation epochs. SG stars are likely enriched with both Na and Eu. Comparing the FG and SG stars, we expect a positive correlation between [Eu/H] and [Na/H]. The observed abundances of M15 do not show this feature.

In another scenario, an $r$-process abundance spread in the natal cloud is the origin of the abundance spread in M15 stars. 
In this case, the number of star formation epochs in M15 would significantly affect the [Na/H]-[Eu/H] relation. If all the stars are formed in a single burst, Eu and Na are independent of each other. Therefore, the distribution would be just random as in the bottom left panel (red colormap and histogram) of Figure~\ref{fig:M15_NaEu}. On the other hand, if M15 stars are formed in multiple star formation bursts, only the FG stars would have Eu spread. The SG star, being formed out of the mixture of stellar winds from the first generation stars, would not show Eu abundance spread. The expected distribution is shown in the bottom-middle panel (green colormap and histogram).

\citet{2009Carretta} investigate the Na-O anticorrelations of 15 GCs and discuss that stars with $\mbox{[Na/Fe]} > \mbox{[Na/Fe]}_\mathrm{min}+0.3$ can be safely regarded as SG stars. Taking $\mbox{[Na/H]} = -2.09$ as the criterion motivated by the lowest Na abundance in W13 stars ($\mbox{[Na/H]} = -2.39$: $\mbox{[Fe/H]} = -2.32$ and $\mbox{[Na/Fe]} = -0.07$), it is clear that the SG stars also show significant [Eu/H] spread comparable to that of the FG stars. Therefore, we conclude that the single star burst is strongly favored as the formation scenario of GC stars.

\subsection{Fraction of GCs with internal Eu spread}
\label{sec:formationrate}
We have seen that a GC can have a significant Eu abundance spread if the last NSM occurred $\lesssim 50$\,Myr before the GC formation. Each NSM creates a region of $\sim 0.02$\,kpc$^{3}$ that allows formation of a GC with Eu spread. Here we estimate how often this happens. 

One NSM creates a region of 0.02\,kpc$^{3}$ for 50\,Myr. Assuming the stellar mass of a galaxy to be $10^7$ \Msun\ at the cosmic age of 1\,Gyr, the average star formation rate is $10^{-2}$\Msun\,yr$^{-1}$. Since a core-collapse supernova (CCSN) occurs in 100\,\Msun\ of stars formed, the CCSN rate is $\sim 10^{2}$\,Myr$^{-1}$. Assuming the NSM rate is $10^{-3}$ of that of CCSN, NSM rate is $\sim 10^{-1}$\,Myr$^{-1}$. Therefore, $\sim 5$ NSMs occur in 50\,Myr, making 0.1\,kpc$^{3}$ of the region with significant Eu spread. In our simulations, most GCs are formed in the central 1\,kpc.\footnote{Although these GCs are formed within $\sim 1\,\mathrm{kpc}$ from the center, most of them are thrown out from the center by galaxy mergers. This might be an important mechanism to sprinkle GCs in MW halo region.} Therefore, assuming the size of the star forming region is $\sim 1$\,kpc$^{3}$, the fraction of GCs with significant Eu spread is 10\,\%. This is consistent with one system in $\sim 7$ metal-poor GCs with $\mbox{[Eu/H]} \lesssim -2$ (M68, NGC6287, NGC6293, M92, NGC6397, M15, and M30 to our knowledge. See also \citealt{Roederer11_GCrprocess, 2011Cohen}). Note that the fraction is highly dependent on the metallicity (or equivalently, the baseline Eu abundance). A GC is more likely to have Eu abundance spread if the homogeneous Eu background is assumed to be low. The estimate on the volume of GC forming cloud bears significant uncertainty. Considering the low metallicities of GCs, 
we argue that the assumption on the formation site may be valid only for early, immature galaxies with low stellar masses of $\lesssim 10^{7} \Msun$. Note that the low [Fe/H] of our target GC does not help realize Eu abundance spread, because we assume a $\mbox{[Eu/H]} = -2.1$ background that is consistent with the metallicity of M15.

\subsection{Origin of {\it r}-process elements}
We assume that NSMs are the origin of {\it r}-process elements. However, other processes such as collapsars \citep{Siegel19_collapsar} and magneto-rotational supernovae \citep{Nishimura15_MRSNe} are also proposed as the origin. Although there is evidence for the delay of the $r$-process in very metal-poor stars \citep{2021Tarumi_VMP_rprocess}, the origin is still under debate.

\begin{figure}[htbp]
	\includegraphics[width=\columnwidth]{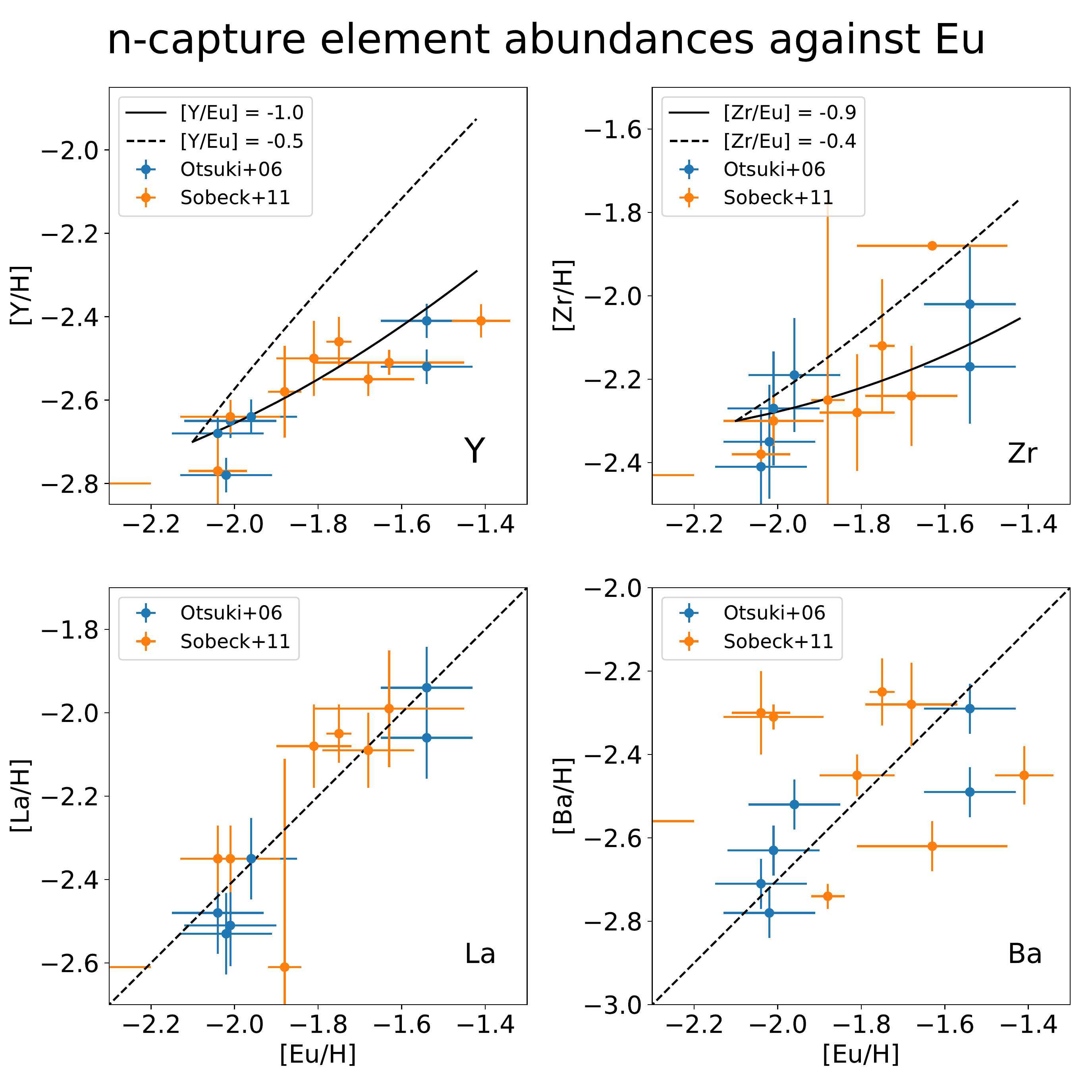}
    \caption{The relations of neutron-capture elements. The top panels show the heavy to light neutron-capture element ratio, and the bottom panels show relations between heavy neutron-capture elements. The dashed lines show the $r$-process contribution of solar $r$-process abundance ratio \citep{2014Bisterzo_sprocess}, and the solid curves show the case of lanthanide-rich abundance ratio \citep{2020Tsujimoto}. The observational data is obtained from \citet{2006Otsuki} and \citet{2011Sobeck}.}
    \label{fig:Eu_others}
\end{figure}

For M15 stars, \citet{2006Otsuki}  point out that the abundances of light $r$-process elements such as yttrium (Y) and zirconium (Zr) show smaller spread compared to that of Eu, as can be found in Figure~\ref{fig:Eu_others}.
There, the lines show the enrichment by  NSMs with two different yields. For the top two panels, we draw the model curve assuming that (i) the homogeneous Y, Zr abundances are $\mbox{[Y/H]} = -2.7, \mbox{[Zr/H]} = -2.3$, (ii) the homogeneous Eu abundance is $\mbox{[Eu/H]} = -2.1$, and (iii) the NSM ejecta has $\mbox{[Y/Eu]} = -1.0, \mbox{[Zr/Eu]} = -0.9$. The last conditions are motivated by the abundance ratio of $r$-II stars \citep{2007Montes_LEPP, 2019Ji_LanthanideFraction, 2020Tsujimoto}. Note that this ratio is lanthanide-rich by $\sim 0.5$\,dex compared to the solar $r$-process abundance. The NSM (or an $r$-process event) that have dominantly enriched M15 have lower [light-$r$/Eu] than the solar $r$-process ratio and is consistent with that of $r$-II stars. Interestingly, 
recent multi-wavelength observations of a kilonova event suggest a low opacity for the ejecta \citep{2018Waxman}. The estimated  lanthanide fraction is also low, but with a significant uncertainty. Future observations and modeling of kilonova will help further clarify the detailed abundances of $r$-process elements produced in an NSM. Different origins for light and heavy $r$-process elements are supported as in previous works (e.g. \citealt{2006Otsuki, 2006Honda_SrRich}). Note that the number of lanthanide-rich $r$-process events that have shaped the Eu spread in M15 is likely one, considering the small physical scale of M15. The lanthanide fraction of the event and $r$-II stars in MW halo are similar, suggesting that the numbers of events that have contributed to the $r$-process element abundances of $r$-II stars are likely one. These stars may have formed in a narrow, isolated environment that are not much affected by lanthanide-poor $r$-process events.

\section{Conclusion}
We have shown with galaxy formation simulation that a GC can have an internal Eu abundance spread by the inhomogeneity of the natal cloud. 
The last $r$-process event should have occured shortly before the GC formation,
and the responsible $r$-process should be lanthanide-rich, as those invoked for the $r$-II stars and Reticulum~II. 
Considering the abundance distribution of Na and Eu, especially the absence of strong correlation, a single star formation epoch model is favored as the formation scenario of GCs, challenging the standard AGB scenario.

\acknowledgments

We thank the anonymous referee for constructive comments. We thank Kenta Hotokezaka, Paz Beniamini, Wako Aoki, and Takuma Suda for fruitful discussions.
Numerical computations were carried out on Cray XC50 at Center for Computational Astrophysics, National Astronomical Observatory of Japan.
Y. T. is supported by JSPS KAKENHI Grant Number 20J21795.

\bibliography{M15}{}
\bibliographystyle{aasjournal}



\end{document}